\RequirePackage{fix-cm}
\documentclass[a4paper,english,aps,prd,twocolumn,10pt,a4paper,superscriptaddress,nofootinbib,nobibnotes,altaffillsymbol,showpacs,showkeys,floatfix]{revtex4-1}

\usepackage[T1]{fontenc}
\usepackage[utf8]{inputenc}
\usepackage{babel}

\usepackage{amsmath}
\usepackage{amssymb}
\usepackage{mathrsfs}
\usepackage{esint} 
\usepackage{bbm} 

\usepackage{graphicx}

\usepackage{tabularx}

\usepackage[unicode=true,
 bookmarks=true,bookmarksnumbered=false,bookmarksopen=true,bookmarksopenlevel=2,
 breaklinks=true,pdfborder={0 0 0},backref=false,colorlinks=true]
 {hyperref}

\hypersetup{pdftitle={Reweighting Lefschetz Thimbles},
 pdfauthor={Stefan Bl\"ucher, Jan M.~Pawlowski, Manuel Scherzer, Mike Schlosser, Sebastian Syrkowski, Felix P.G.~Ziegler},
 linkcolor=BlueViolet,anchorcolor=BlueViolet,citecolor=BlueViolet,filecolor=BlueViolet,urlcolor=BlueViolet}

\allowdisplaybreaks
\usepackage{color}
\usepackage[usenames,dvipsnames]{xcolor}

\usepackage[normalem]{ulem}

\def\eq#1{(\ref{#1})}

\def\s0#1#2{\mbox{\small{$ \frac{#1}{#2} $}}}
\def\0#1#2{\frac{#1}{#2}}

\newcommand{\be}{\begin{equation}}
\newcommand{\ee}{\end{equation}}
\newcommand{\bear}{\begin{eqnarray}}
\newcommand{\eear}{\end{eqnarray}}
\begin{document}

\title{Reweighting Lefschetz Thimbles}

\author{Stefan Bl\"ucher}

\affiliation{Institut für Theoretische Physik, Universität Heidelberg, Philosophenweg
16, 69120 Heidelberg, Germany}

\author{Jan M.~Pawlowski}

\affiliation{Institut für Theoretische Physik, Universität Heidelberg, Philosophenweg
16, 69120 Heidelberg, Germany}

\affiliation{ExtreMe Matter Institute EMMI, GSI, Planckstraße 1, D-64291 Darmstadt,
Germany}

\author{Manuel Scherzer}

\affiliation{Institut für Theoretische Physik, Universität Heidelberg, Philosophenweg
16, 69120 Heidelberg, Germany}

\author{Mike Schlosser}

\affiliation{Institut für Theoretische Physik, Universität Heidelberg, Philosophenweg
16, 69120 Heidelberg, Germany}

\author{Ion-Olimpiu Stamatescu}

\affiliation{Institut für Theoretische Physik, Universität Heidelberg, Philosophenweg
16, 69120 Heidelberg, Germany}

\author{Sebastian Syrkowski}

\affiliation{Institut für Theoretische Physik, Universität Heidelberg, Philosophenweg
16, 69120 Heidelberg, Germany}

\author{Felix P.G.~Ziegler}

\affiliation{Institut für Theoretische Physik, Universität Heidelberg, Philosophenweg
16, 69120 Heidelberg, Germany}

\keywords{sign problem, complex Langevin evolution, Lefschetz thimbles}

\begin{abstract}
  One of the main challenges in simulations on Lefschetz thimbles is
  the computation of the relative weights of contributing thimbles. In
  this paper we propose a solution to that problem by means of
  computing those weights using a reweighting procedure. Besides we
  present recipes for finding parametrizations of thimbles and
  anti-thimbles for a given theory.  Moreover, we study some
  approaches to combine the Lefschetz thimble method with the Complex
  Langevin evolution.  Our numerical investigations are carried out by
  using toy models among which we consider a one-site $z^4$ model as
  well as a $U(1)$ one-link model.
\end{abstract}
\maketitle


\section{Introduction}
\label{sec:intro}
QCD at vanishing and finite temperature is one of the best tested
theories in high energy physics.  At the present moment theoretical
predictions from first principle lattice simulations match remarkably
well with experimental data for instance from heavy ion collisions, see
e.g.~\cite{Bellwied:2015rza, Bazavov:2014xya, Akiba:2015jwa}.
However, at finite chemical potential lattice simulations suffer from
the sign problem, which a priori prohibits simulations based on
importance sampling.  This severely limits the access to the largest
part of the QCD phase diagram. By now, there are many approaches
towards a solution of the sign problem.  Overviews addressing
developments in finite density QCD over the last years can e.g.~be
found in \cite{deForcrand:2010ys, Aarts:2013lcm, Sexty:2014dxa,
  Gattringer:2014nxa, Scorzato:2015qts}.  Amongst those are Taylor
expansions \cite{Allton:2002zi}, simulations at imaginary chemical
potential \cite{deForcrand:2002hgr, DElia:2002tig}, reweighting
\cite{Fodor:2001au}, the density of states method
\cite{Langfeld:2014nta}, dual formulations \cite{Gattringer:2014nxa},
the Complex Langevin method \cite{Aarts:2008rr, Sexty:2013ica} and the
Lefschetz thimble method \cite{Witten:2010cx, Cristoforetti:2012su}.
So far none of those methods have been able to give reliable results
for $\mu/T\gtrsim 1$.  In this work we focus on the Lefschetz thimble
approach.  In Euclidean space-time the Lefschetz thimble approach has
been applied to bosonic theories as well as to (low-dimensional) QCD
in \cite{Cristoforetti:2012su, Cristoforetti:2013wha,
  Cristoforetti:2014gsa, DiRenzo:2017igr, DiRenzo:2017omx,
  Schmidt:2017gvu}.  Recent applications to fermionic theories (such
as the Thirring model) can be found in \cite{Alexandru:2015xva,
  Kanazawa:2014qma}. Moreover, field theories in Minkowski space-time
formulated on the Schwinger-Keldysh contour have been studied using
the thimble formalism in \cite{Alexandru:2016gsd}. Algorithmic
improvements to the holomorphic gradient flow method were proposed in
\cite{Tanizaki:2017yow}. Recent contributions in the field more generally involve 
complex manifolds close to Lefschetz thimbles 
that are optimized such that they ameliorate the sign 
problem, see e.g.~\cite{Mori:2017pne, Mori:2017nwj, Alexandru:2018fqp, Bursa:2018ykf}.

 As we see later, the CLE and the Lefschfetz
thimbles are closely related. Studies investigating the interplay and
the connection between the two approaches can be found in
\cite{Aarts:2014nxa, Nishimura:2017vav}.

The Lefschetz thimble method relies on a deformation of the
integration path. By construction the imaginary part of the action is
constant on the transformed paths, which are called Lefschetz
thimbles. There are two basic algorithmic frameworks
\cite{Cristoforetti:2012su, Alexandru:2015xva} providing recipes for
Monte Carlo simulations on the Lefschetz thimbles. The first employs
Monte Carlo simulations directly on the thimbles. The latter continuously
deforms the original integration path close to the actual thimbles to
lessen the sign problem. 

In this work we address a few key challenges to the the Lefschetz
thimble method and propose algorithmic improvements. One of the main
problems with Monte Carlo simulations on Lefschetz thimbles is to
determine the weights of the thimbles relative to each other. This
difficulty arises as the original path integral is decomposed into a
sum of integrals over multiple thimbles. We show that this difficulty
can be overcome by a standard Monte Carlo determination of the ratios
of the real partition functions on the thimbles. This is facilitated
by a novel reweighting procedure which is generally applicable to
field theories. In this work we assume prior knowledge on a parametrization of the
contributing thimbles. To find this parametrization we propose two
algorithms which can be generalized to higher dimensional theories.
However, the reweighting procedure does not rely on knowing a
parametrization. Our ideas are put to work in simple models, i.e.~ordinary
integrals. Among those we consider a one-site quartic model with a
$\frac{\lambda}{4}z^4$ term as well as a $U(1)$ one-link model. 

The paper is organized as follows. We start by briefly revisiting the
idea behind Lefschetz thimbles, see Sec.~\ref{sec:method_intro}. In
Sec.~\ref{sec:search} we propose two algorithms to find thimbles and
their parametrizations necessary for Monte Carlo integration.  In
Sec.~\ref{sec:monte_carlo} we present our idea
of sampling on multiple thimbles taking into account the relative
weights of different thimbles. Sec.~\ref{sec:results} introduces the
toy models we use for numerical investigations together with our
results. We conclude this paper in Sec.~\ref{sec:concl}.  During the
research for this paper we have also developed many ideas to combine
the Complex Langevin evolution and the Lefschetz Thimble method. While
none of those approaches lead to generally applicable algorithms, they
still provide some useful insight into the structure of the models,
hence we give some of those ideas and corresponding results in
App.~\ref{app:LCOOL}.

\section{The Lefschetz thimble method}
\label{sec:method_intro}
In the following we briefly revisit the Lefschetz thimble method.  The
idea behind the approach is to rewrite the path integral measure over
a real manifold\cite{Witten:2010cx} to circumvent the sign problem by
allowing Monte Carlo sampling on this
manifold\cite{Cristoforetti:2012su}.  Here we explain the Lefschetz
thimble method by using the example of simple one-dimensional
integrals.  Consider a complex action of a real variable $S(x)$. Next,
extend the real axis to the complex plane
$\mathbb{R}\rightarrow\mathbb{C}$, i.e. $x \to z = x + i y$.  Given
the stationary points $z_{\sigma}$ of $S(z)$
\begin{equation} 
\left. \frac{\partial S}{\partial z}\right|_{z=z_{\sigma}}=0\,,
\end{equation}
one can define a real path in the complex plane
$D_{\sigma}\subset \mathbb{C}$ as the solution of the steepest descent
equation ending at $z_{\sigma}$
\begin{equation}
\frac{\partial}{\partial \tau}z=-\overline{\frac{\partial S}{\partial z}}\,,
\end{equation}
this path is called a Lefschetz thimble.  The action has constant
imaginary part along the thimble. The integral can be decomposed into
integrals over all $D_{\sigma}$
\begin{align}
  \nonumber Z=&\int_{D} dz\,e^{-S}=\sum_{\sigma}n_{\sigma}
                e^{-i\text{Im}\left[S\left(z_{\sigma}\right)\right]}\int_{D_{\sigma}}dz\, 
                e^{-\text{Re}\left[S\left(z\right)\right]} \\[1ex]
  \equiv&\sum_{\sigma}n_{\sigma}e^{-i\text{Im}\left[S\left(z_{\sigma}\right)\right]} Z_{\sigma}\,,
\label{eq:Thimble_basics}
\end{align}
where $D\subset \mathbb{R}$ is the original real domain and
$n_{\sigma}$ is the intersection number of the steepest ascent path
(unstable thimble) with the original domain $D$.  One can now
formulate Monte Carlo algorithms based on
\eqref{eq:Thimble_basics}, see e.g.~\cite{Cristoforetti:2012su,
  Alexandru:2015xva, Nishimura:2017vav}.  Observables are then
computed in the usual way
\begin{align}
\nonumber\left<\mathcal{O}\right>&=\frac{1}{Z}\sum_{\sigma}n_{\sigma}
e^{-i\text{Im}\left[S\left(z_{\sigma}\right)\right]}\int_{D_{\sigma}}dz\,\mathcal{O}e^{-\text{Re}\left[S\left(z\right)\right]}\\[1ex]
\nonumber&=\frac{1}{Z}\sum_{\sigma}n_{\sigma}e^{-i\text{Im}\left[S
\left(z_{\sigma}\right)\right]}Z_{\sigma}\left<\mathcal{O}\right>_{\sigma}\\[1ex]
&=\frac{\sum_{\sigma} n_{\sigma} e^{-i\text{Im}S\left(z_{\sigma}\right)}Z_{\sigma} 
\left<\mathcal{O}\right>_{\sigma}}{\sum_{\sigma} n_{\sigma} e^{-i\text{Im}S\left(z_{\sigma}\right)}Z_{\sigma}}\,,
\end{align}
with the only difference, that it has to be computed on every thimble.
There are two practical problems with this approach: 
\begin{enumerate}
\item Finding all contributing thimbles can be a challenging task.
\item For the case of multiple contributing thimbles there is no
  simple way so far to access the relative weights,
  i.e. the ratio $Z_{\sigma_{i}}/Z_{\sigma_{j}}$.
\end{enumerate} 
Both are challenging problems without a general solution so far. An
approach to the first problem is the holomorphic gradient flow, which
approximates the thimble structure, by simulating on a deformation of
the original domain close to thimbles \cite{Alexandru:2015xva}. In
\cite{DiRenzo:2017omx} the second problem regarding relative weights
has been approached by using known results in some parameter regions.

In the following we propose general solutions to both of those
problems which do not rely on approximations. In this work we
demonstrate our solutions by means of simple models.

\section{Finding thimbles}

In this section we propose two algorithms which can be used to
systematically find contributing thimbles. This is put to work in
simple one-dimensional integrals. Generalizations to higher dimensions
might be expensive. The first algorithm scans the real axis in search
of intersecting anti-thimbles, while the second algorithm projects
points in the complex plane onto thimbles, in order to determine a
numerical parametrization of the thimbles. Both algorithms also apply
in higher dimensions, however the numerical costs may rise
exponentially with the number of lattice points. This is currently
investigated for gauge theory in \cite{HD-Bielefeld}.

\label{sec:search}
\subsection{Axis scan}
\label{sec:scanaxis}
\begin{figure}[t]
{
\centering
\includegraphics[width=0.9\columnwidth]{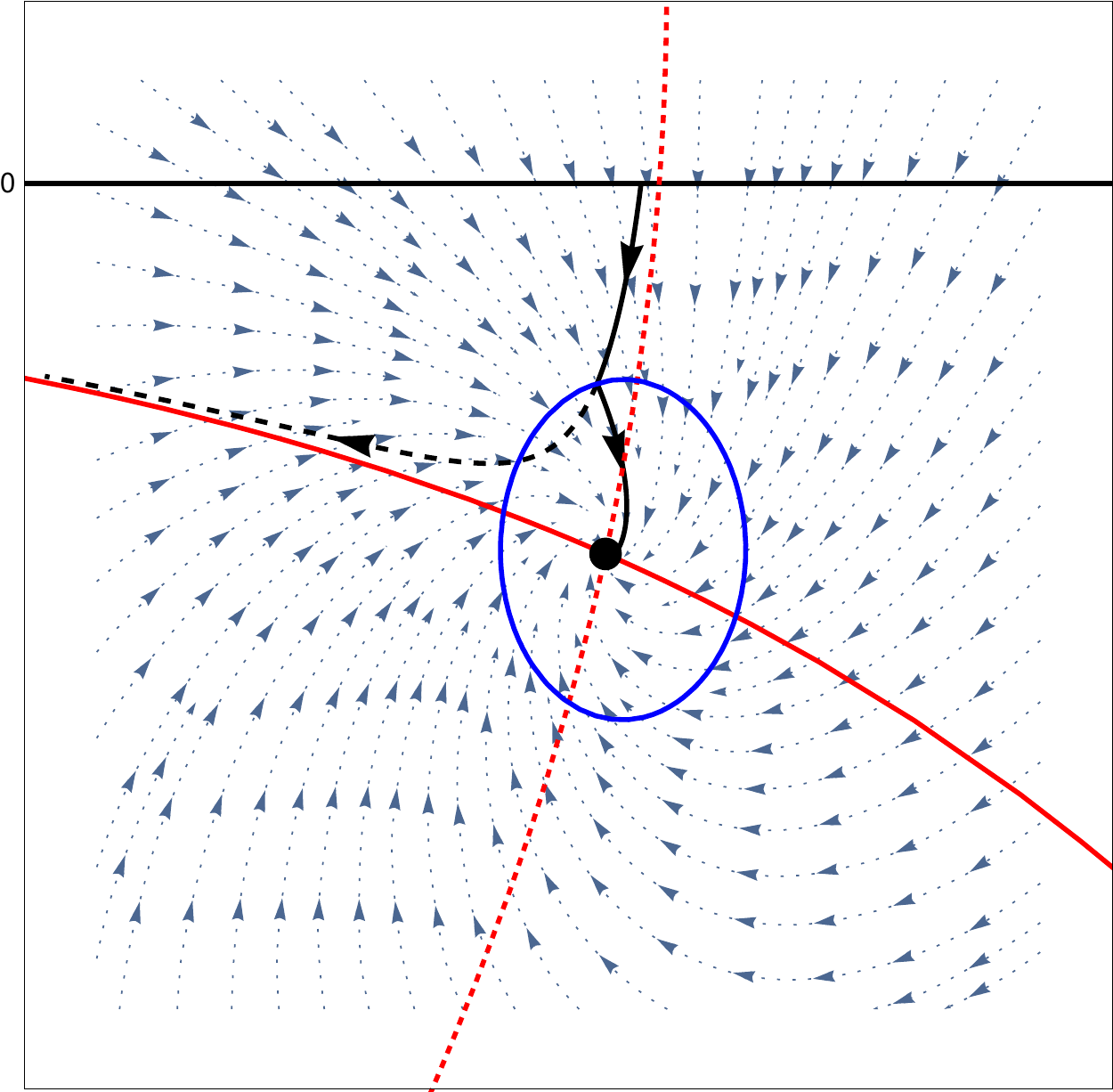}
}

\caption{Visualization of the fixed point search via an axis scan. Red
  are the thimble (solid) and the anti-thimble (dashed). The steepest
  ascent equation \eqref{eq:norm_sd} is solved using a starting point
  on the real axis close to the anti-thimble, once it is close to the
  fixed point (i.e. the derivative of the action is smaller than some
  value $\delta$, visualized by the blue circle), the flow is switched
  according to equation \eqref{eq:searchflows}, and will end in the
  fixed point (solid black arrows). The steepest ascent without
  switching close to the fixed point will asymptotically approach the
  thimble (black dashed arrow). }
\label{fig:search}
\end{figure}
Since the only contributing thimbles are those with non-zero
intersection number of the anti-thimble with the original manifold,
one can find all contributing fixed points by scanning the manifold
for such intersections.  This can be a challenging problem in higher
dimensional theories, however importance sampling by Monte Carlo
methods in parameter regions without a sign problem or in the phase
quenched theory might give good starting points for such searches. In
the following we describe the searching algorithm for the case of
simple integrals, i.e.~the original manifold is an interval
$\left[a,b\right]\in\mathbb{R}$. The algorithm is the following
\cite{Schlosser}, 
\begin{enumerate}
\item Choose a starting point on the real axis.
\item Solve the steepest ascent equation 
\begin{equation}
\frac{\partial z}{\partial \tau}=\overline{\frac{\partial S}{
\partial z}}\left/\left|\overline{\frac{\partial S}{\partial z}}\right|\right.\, ,
\label{eq:norm_sd}
\end{equation}
using the starting point as an initial condition.
\item If the derivative of the action becomes small
\begin{equation}
\left|\frac{\partial S}{\partial z}\right| < \delta\, ,
\end{equation}
the flow is close to a fixed point of the action.
\item Depending on the structure of the fixed point, one can now reach
  it by looking at the Langevin flow (LF)
\begin{equation}
\dot{z}=-\frac{\partial S}{\partial z}\,,
\end{equation} 
and changing the sign according to the following prescription
\begin{align}
   \dot{z} =
   \begin{cases}
     -\frac{\partial S}{\partial z} & \text{FP attractive under LF }  \\[1ex]
     +\frac{\partial S}{\partial z} & \text{FP repulsive under LF} \\[1ex]
      \pm e^{i\pi/2}\frac{\partial S}{\partial z} & \text{FP circular under LF} 
   \end{cases}.
\label{eq:searchflows}
\end{align}
All those cases have to be tested, and one of them will end in the fixed point.
\end{enumerate}
This algorithm is visualized in Fig. \ref{fig:search}.

Once the fixed points are known, the numerical parametrization of the
thimbles can be computed. In the case of one dimensional integrals,
this boils down to solving one dimensional differential equations. We
do so by solving the normalized steepest descent equation
\begin{equation}
\frac{\partial z}{\partial \tau}=-\overline{\frac{\partial S}{\partial z}}\left/\left|\overline{
\frac{\partial S}{\partial z}}\right|\right.\, ,
\label{eq:steep_des_asc}
\end{equation}
with opposite sign starting close to the fixed point
\cite{Aarts:2014nxa}. The reason for the normalization with the
absolute value will become clear later, however it also helps with
numerical stability when solving the steepest descent equation. Note
that this normalization is simply a rescaling of the flow parameter
$\tau$.

\subsection{Thimble cooling}
\label{sec:cooling}

\begin{figure}[t]
   {
	\centering
	\includegraphics[width=.45\columnwidth]{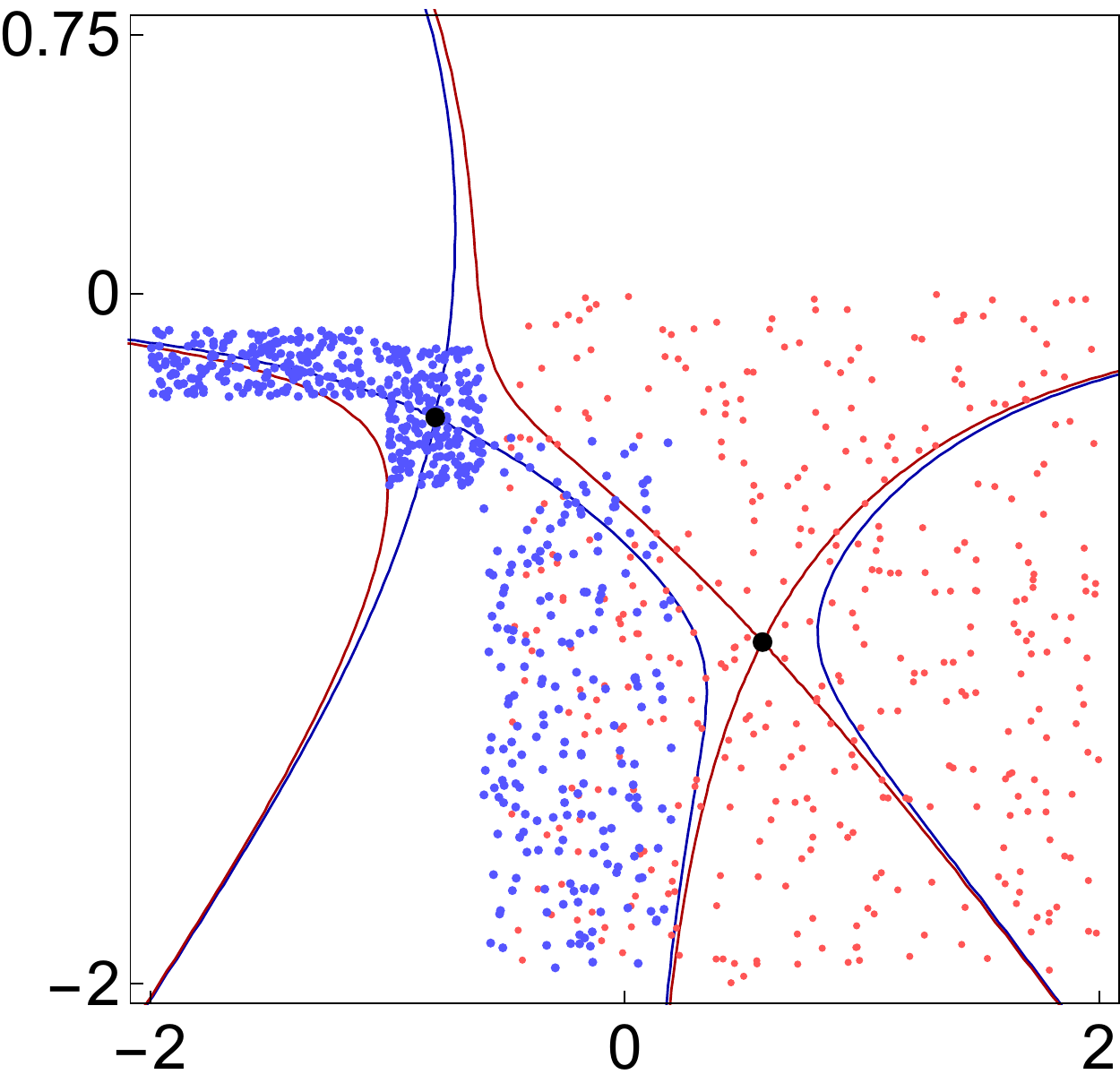}~
	\includegraphics[width=.45\columnwidth]{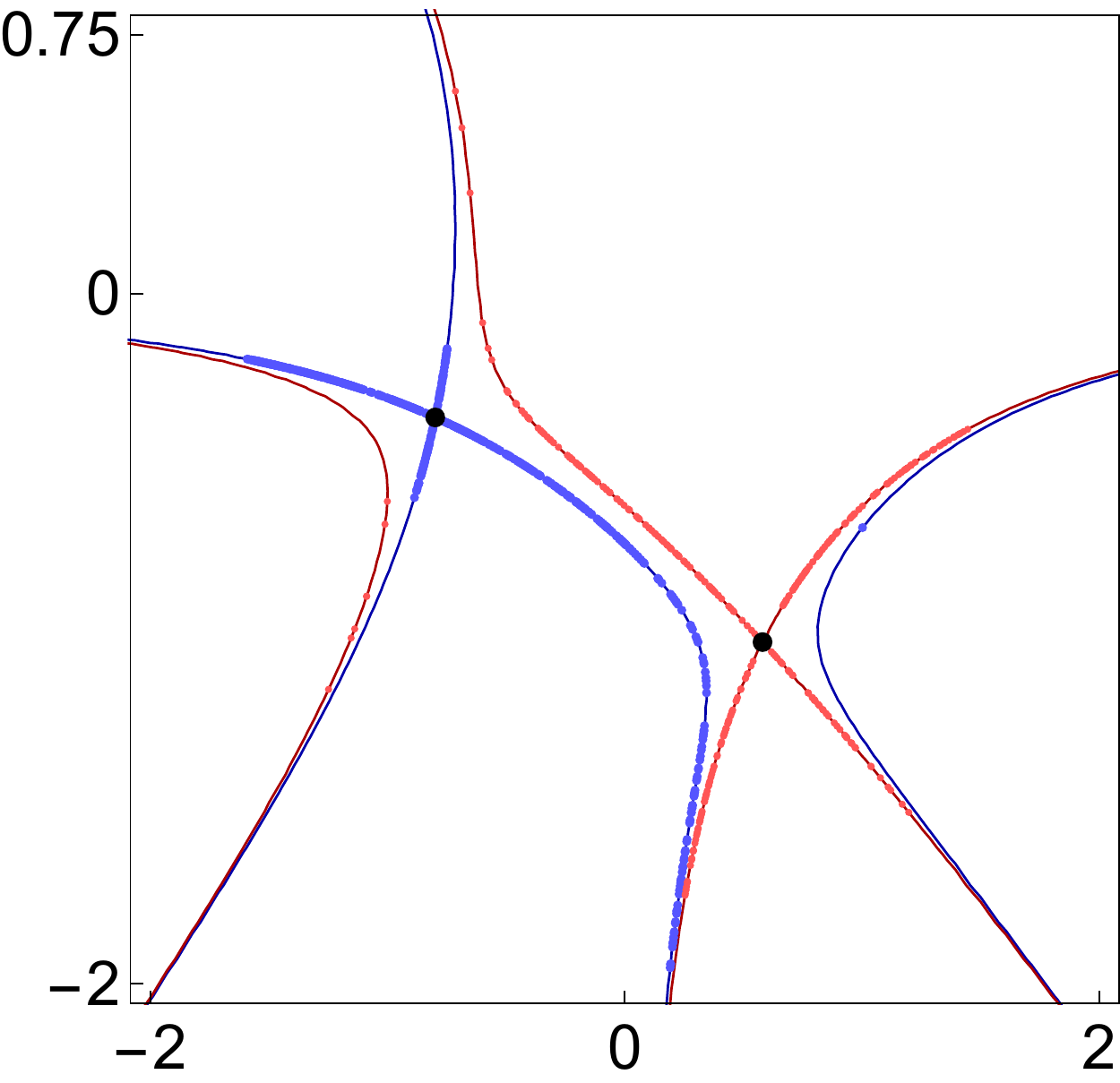}
	}
	\caption{ Complex plane with the solutions to the gradient
          equations (\ref{eq:GradFlow}) for a given grid of initial
          conditions for the left thimble (blue) and the right thimble
          (red). In the left plot we show a possible choice of initial
          conditions for the gradient equations. The right plot
          depicts the resulting distribution of points lying on the
          thimble and anti-thimble. Moreover other curves not passing
          through the fixed point on which the imaginary part of the
          action is constant are hit by the algorithm. }
    \label{fig:CoolingSearch}
\end{figure}

In this section we propose a straight-forward algorithm to find
parametrizations of all thimbles and anti-thimbles for a given action
where the only required information is the knowledge about the fixed
points. We show results for the $z^4$ model with parameters
$\sigma = 1, \lambda = 1, h = 1 + i$. For a definition of this model
see Sec.~\ref{sec:modelz4}. The idea is to minimize the distance of
any set of points in the complex plane to curves on which the
imaginary part of the action is constant. This leads to the following
definition.
\begin{equation}
M_{\sigma}(x, y) = | \mathrm{Im} S(x + i y) - \mathrm{Im} S(z_{\sigma})|^2\,,
\label{eq:CoolFunc}
\end{equation}
where $\sigma$ labels the stationary points. We call $M_{\sigma}(x, y)$ the cooling function.
The numerical minimization procedure is facilitated by the following gradient equations
\begin{align}\nonumber 
\dot x &= - \frac{\partial M_{\sigma}(x, y)}{\partial x}\,,  \\[1ex]
\dot y &= - \frac{\partial M_{\sigma}(x, y)}{\partial y}\,. 
\label{eq:GradFlow}
\end{align}
By construction the gradient equations orthogonally project a given
point on a curve with constant imaginary part of the action.  As
initial conditions for (\ref{eq:GradFlow}) we use a grid of points in
the complex plane.  This is shown in the left plot in
Fig.~\ref{fig:CoolingSearch}. The choice of a random grid is
arbitrary.  We could have also chosen a regular grid.  In the right
plot of Fig.~\ref{fig:CoolingSearch} we show the result of the
solution to the gradient equations. The procedure works well for a
large box of initial conditions (red) as shown for the right fixed
point (black). The resulting set of points lies on the thimble, the
anti-thimble and an additional curve on the left without physical
relevance.  Alternatively, we can start with a small rectangle around
the fixed point, see the blue points in the left plot. From the
resulting points flowed to thimble and anti-thimble we can choose the
next set of initial conditions along e.g.~the thimble and repeat the
procedure iteratively. Note that also from looking at the flow lines
of (\ref{eq:GradFlow}) we can determine suitable areas for initial
conditions.

This method provides a useful tool to find parametrizations to the
thimbles (and anti-thimbles) by interpolating the flowed set of
points. Moreover, by knowing each anti-thimble we can map out if it
intersects with the original integration manifold thus enabling us to
determine whether and how much the corresponding thimble contributes.
In particular, the method could be applied in higher dimensional
theories where the minimization procedure is combined with importance
sampling around the fixed point. From there thimbles and anti-thimbles
can be successively parametrized as illustrated in blue in
Fig.~\ref{fig:CoolingSearch}.

Thimble cooling has the potential advantage over the
procedure in Sec.~\ref{sec:scanaxis} that one does not have to solve
the holomorphic gradient flow in many directions but one directly
flows to the (anti-)thimbles.  The cooling method put forward here can
also be used to reduce numerical discretizations artifacts (see
Sec.~\ref{sec:results}) in the thimble parametrization. 

We remark, that a generalization of thimble cooling to higher
dimensional integrals may in general prove difficult due to the dimensionality of
the hyper-surface parametrized by
$\mathrm{Im}[S(z)] = \mathrm{const}$.  In App.~\ref{app:advLC} we
propose a combination of Lefschetz thimble and complex Langevin, which
samples around all thimbles. This can also be used as a starting point
for thimble cooling.

\section{Monte Carlo simulations on Lefschetz thimbles}
The previous section dealt with finding parametrizations for the
thimbles. In this section we propose an algorithm for simulating on
thimbles provided its parametrization is known. We also show how to
compute the ratio of partition functions from within Monte Carlo
simulations.
\label{sec:monte_carlo}

\subsection{Reweighting on thimbles}

Once the parametrization of the thimble is known, we can simply
rewrite the partition function $Z_{\sigma}$ on the thimble as
\begin{equation}
  \int_{D_{\sigma}} dz\,e^{-\text{Re}\left[S(z)\right]} = \int_{a}^{b}  d\tau\,
  e^{-\text{Re}\left[S_\sigma(\tau)\right]} J_\sigma(z(\tau))\, ,
\label{eq:jacobian}
\end{equation}
where $S_{\sigma}(\tau)=S(z(\tau))$ is the action evaluated on the
thimble $D_{\sigma}$.  We have also rewritten the integral to run over
the flow parameter in \eqref{eq:steep_des_asc} and introduced integral
boundaries, which are defined by the domain of $\tau$. This introduces
the complex Jacobian $ J_{\sigma} = \partial z / \partial \tau$ on
$D_\sigma$. For the right hand side of \eq{eq:jacobian} we can apply a
real Langevin simulation or Monte Carlo sampling along the
thimble. The Jacobian is dealt with via reweighting,
\begin{equation}
  \left< \mathcal{O}\right> =\frac{\left<\mathcal{O}\,J_\sigma\right>}{
    \left<J_\sigma\right>}\,.
\label{eq:reweight}
\end{equation}
Monte Carlo sampling now produces samples according to the distribution
\begin{equation}
p_i(\tau_i)= e^{-\text{Re}\left[S(\tau_i)\right]}\,.
\label{eq:distribution}
\end{equation}
So far we have dealt with a single thimble. The reweighting equation
\eqref{eq:reweight} for multiple thimbles becomes
\begin{equation}
\left<\mathcal{O}\right>=\frac{\sum_{\sigma} n_{\sigma} 
e^{-i\text{Im}\left[S\left(z_\sigma\right)\right]}Z^{r}_{\sigma} \left<
\mathcal{O}\,J_\sigma \right>_{\sigma}^{r}}{\sum_{\sigma} n_{\sigma} 
e^{-i\text{Im}\left[S\left(z_\sigma\right)\right]}Z^{r}_{\sigma} \left<J_\sigma\right>_{\sigma}^{r}}\,,
\label{eq:obs_full}
\end{equation}
where we have defined 
\begin{equation}
\langle \mathcal{O} \rangle_{\sigma}^{r} = 
\frac{1}{Z_{\sigma}^r}\int_{a}^{b} d\tau\,e^{-\text{Re}\left[S_\sigma\left(\tau\right) \right]} \mathcal{O}\, ,
\end{equation}
with 
\begin{equation}
Z_{\sigma}^{r}= \int_{a}^{b} d\tau\,e^{-\text{Re}\left[S_\sigma\left(\tau\right)\right]}\,.
\end{equation}

Note that in (\ref{eq:obs_full}) the thimbles are weighted with their partition functions 
which have to be determined within the simulation.
\subsection{Computing the partition function weights}
\label{sec:weights}
Now that we have a simple algorithm for computing observables on the thimbles, we can proceed to the problem of how to compute the weights. With the above definition of $S_{\sigma}$ and considering only two thimbles for simplicity, we look at the ratio of their partition functions, i.e. we choose one thimble as a ``master'' thimble and divide the numerator and denominator of \eqref{eq:obs_full} by its partition function.  
The following identity states the ratio of partition functions 
\begin{equation}
\frac{Z_{1}^r}{Z_{2}^r}= 
\left<e^{\text{Re}\left[S_2-S_1\right]}\right>_{2}^{r}\,,
\label{eq:ratio}
\end{equation}
provided (i) the integrals over the thimbles have the same boundaries
and (ii) the flow parameters $\tau$ on both thimbles can be identified
-- if the latter does not hold, an additional Jacobian must be taken
into account. For a derivation of \eqref{eq:ratio} see
App.~\ref{sec:deriv}. (i) can be enforced by using suitable variable
transformations, see App.~\ref{app:integration}, while (ii) is
guaranteed by normalizing the steepest descent equations, as we did in
\eqref{eq:norm_sd}.  Hence, it is possible to compute the ratio during
the Monte Carlo simulation, which is necessary in higher dimensional
integrals, e.g.~in field theories.

\section{Applications}
\label{sec:results}
We investigate different models with varying complexity to test our
algorithms. First we look at a model with only one contributing
thimble, which is a good test case for setting up the Monte Carlo
simulation. Next we address a model with two contributing thimbles,
whose flow parameters run over the same interval
$\tau\in\left[-\infty,\infty\right]$. Finally we investigate the U(1)
one link model, which is a model for a simple gauge theory with
fermions and has thimbles that end in poles. This model is quite
general in the sense that it contains all features that are to be
expected in more complicated cases such as field theories.

\subsection{One-site $z^4$ model}
\label{sec:modelz4}
\begin{figure}[t]
\includegraphics[width=0.46\columnwidth]{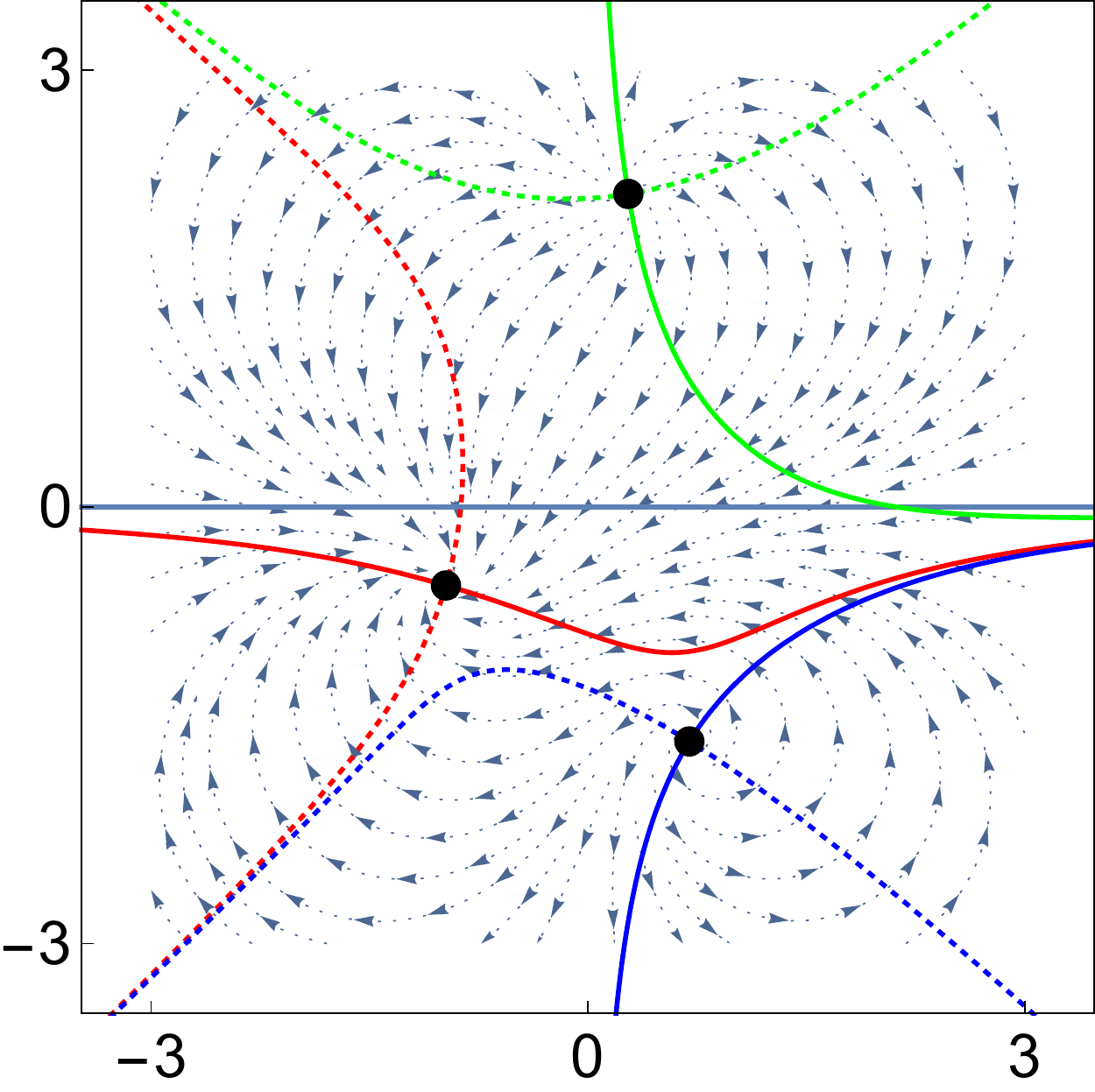}~
\includegraphics[width=0.46\columnwidth]{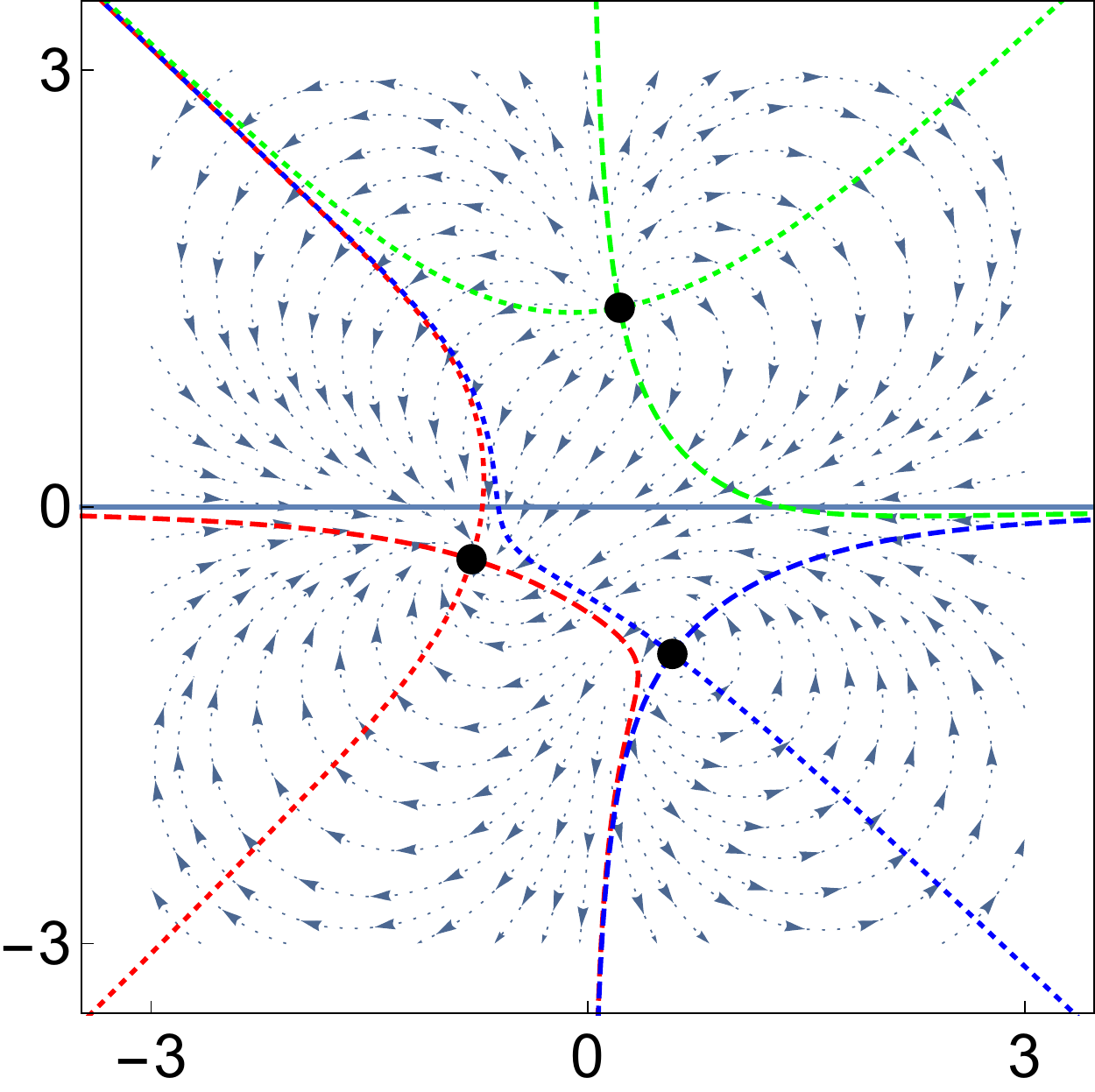}

\caption{Complex plane with drift and thimble structure of the $z^4$
  one-site model with one (left) and two (right) contributing
  thimbles.}
\label{fig:z4}
\end{figure}

The one-site $z^4$ model generically consists of three thimbles, which end in different asymptotic
regions at infinity. This structure makes the model a rather simple
test case for the algorithms we propose.  The model is given by the
action
\begin{equation}
S(z)=\frac{\sigma}{2} z^2 + \frac{\lambda}{4} z^4 + hz\,, 
\end{equation}
for more details see e.g.~\cite{Aarts:2014nxa}.  We can choose the
models parameters such that there are one or two contributing
thimbles, i.e. with $n_{\sigma}\neq 0$:
\begin{enumerate}
\item For $\sigma=1$, $\lambda=1/3$ and $h=1+i$ there is only one
  contributing thimble.
\item For $\sigma=1$, $\lambda=1$ and $h=1+i$ there are two contributing thimbles.
\end{enumerate}
Both cases are shown in Fig. \ref{fig:z4}.

\begin{figure}[t]
{
\centering
\includegraphics[width=0.9\columnwidth]{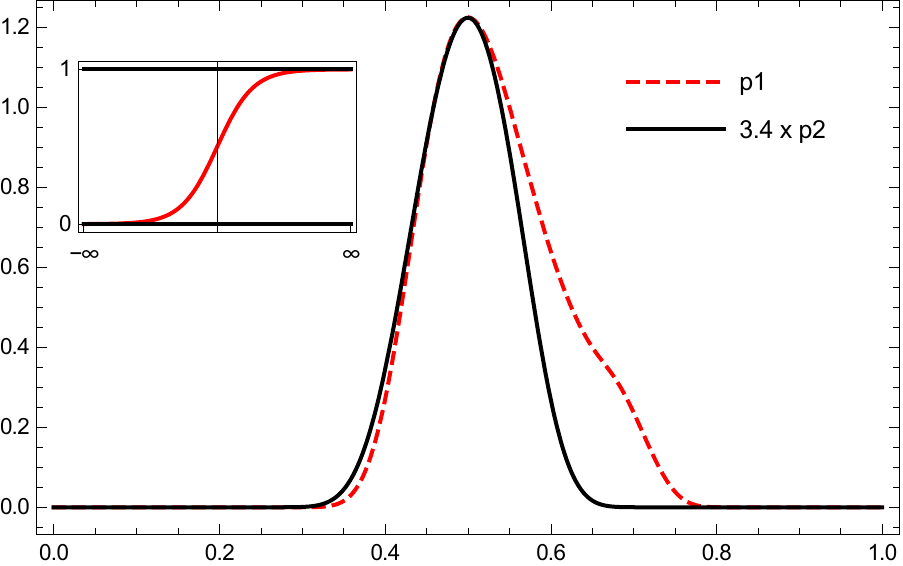}
}

\caption{Boltzmann factor $\text{exp}(-\text{Re}(S))$ vs. flow
  parameter $\tau$ on the contributing thimbles in the $z^4$
  model. The integration ranges have been mapped to $\left[0,1\right]$
  and the transformation is shown in the inlay, for details see
  App.~\ref{app:integration}. Here we choose the free parameter in the
  transformation \eqref{eq:trafo3} to be $\xi=0.25$}
\label{fig:dist_z4}
\end{figure}

The distributions $\text{exp}\left(-\text{Re}\left(S\right)\right)$ on
both thimbles are shown in Fig.~\ref{fig:dist_z4}, where they have
been mapped onto the interval $\left[0,1\right]$, see 
App.~\ref{app:integration}. Note that this is not necessary for the
simulation in this model, since the original domains already
overlap. There we see that both distributions fall off exponentially
which guarantees numerically stable simulations. Numerical results for
the observable $\left<z^{2}\right>$ as well as the ratio of partition
functions from \eqref{eq:ratio} are given in Table
\ref{tab:results_z4}, those simulations have been performed without
employing the variable transformation for simplicity. For the
simulations we have collected $\O(10^{10})$ data points. Errors have
been estimated via a standard Jackknife analysis.

\begin{table}[t]
\def\arraystretch{1.5}
\begin{tabular}{ |c | c | c |}
  \hline			
  $\mathcal{O}$ & numerical & exact \\
\hline \hline
$z^4$--1 thimble &  &  \\
\hline

  Re$z^{2}$ & $0.73922(6)$& $0.73922 $ \\
  Im$z^{2}$ & $0.63006(4) $& $0.630089 $ \\

\hline \hline
$z^4$--2 thimbles &   &  \\
\hline
  Re$z^{2}$ & $0.509299(5) $  & $0.509297$\\
  Im$z^{2}$ & $0.305819(3) $  & $ 0.305815$ \\
  \hline
  $\left.Z_2/Z_1\right|_{T_1}$ &$0.2253778(4)$ & $0.2253779$ \\ 
   $\left.Z_1/Z_2\right|_{T_2}$ &$4.436(12)$ & $4.437$ \\
\hline
\end{tabular}
\caption{Numerical results and exact values of observables with statistical errors. Possible deviations are caused by numerical discretization errors.}
\label{tab:results_z4}
\end{table}

\subsection{U(1) one link model}
\label{sec:modelU1}
\begin{figure}[h]
{
\centering
\includegraphics[width=0.9\columnwidth]{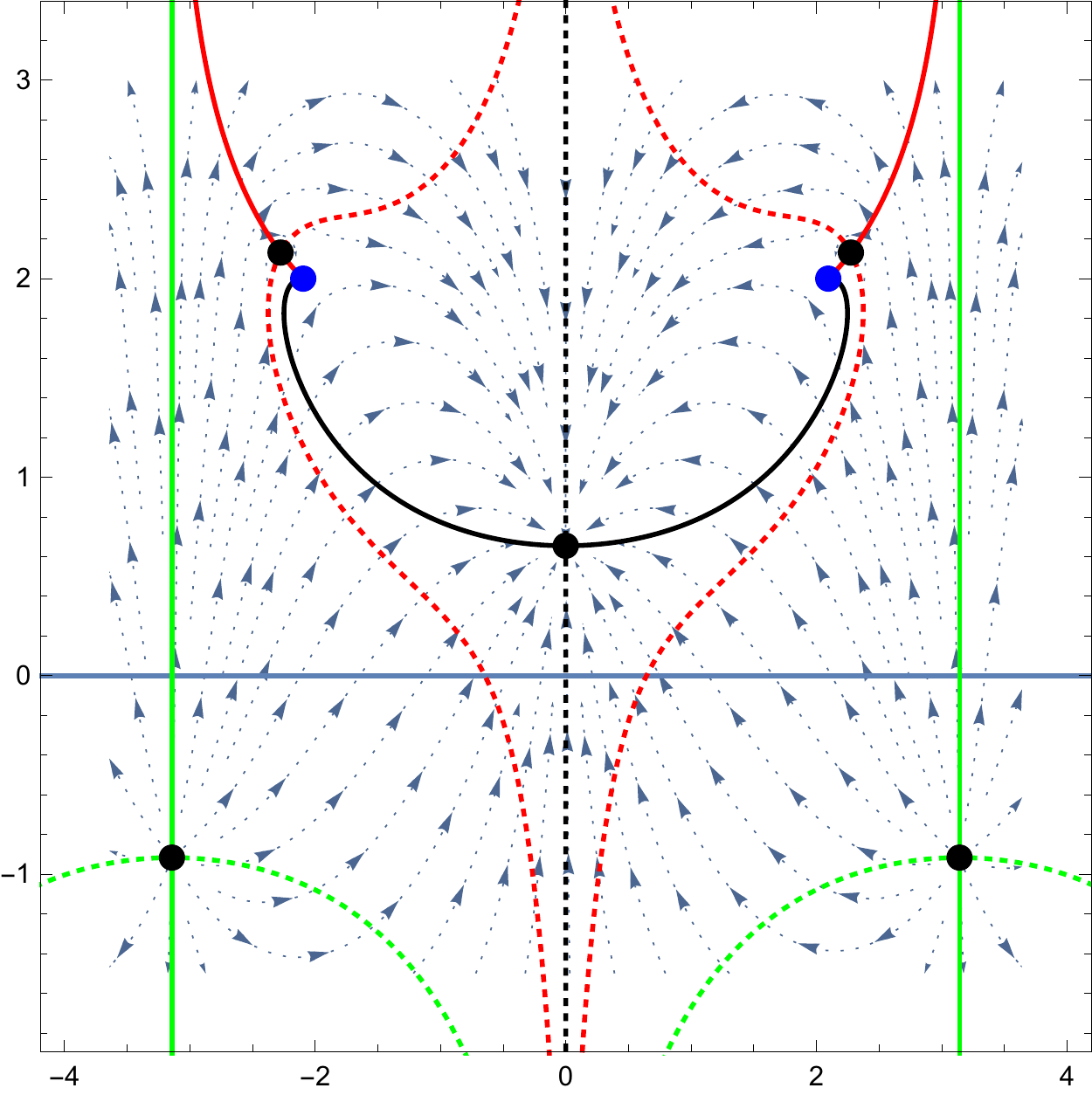}
}
\caption{Complex plane with thimble structure and drift of the U(1)
  one link model. Note that due to periodicity the green thimbles
  (full vertical lines) which are not contributing are actually the
  same.}
\label{fig:U1}
\end{figure}

After having demonstrated the viability of our methods in a simple model, we look at a more difficult model, namely the U(1) one link model with a finite chemical potential $\mu$.
In addition to having multiple contributing thimbles, this model has thimbles that end in poles at finite values of the flow time making it necessary to transform the integrals such that the the distributions on the thimbles overlap. This model is a suitable testbed for our methods, since it contains general features that will also be present in more realistic theories.
The model's action reads \begin{equation}
S(x)=-\beta \text{cos}(x)-\text{log}\left(1+\kappa \text{cos}(z-i\mu)\right)\,,
\end{equation}

where $\kappa=2$, $\beta=1$ and $\mu=2$. Its thimble structure is shown in Fig.~\ref{fig:U1}.
\begin{figure}
{
\centering
\includegraphics[width=0.9\columnwidth]{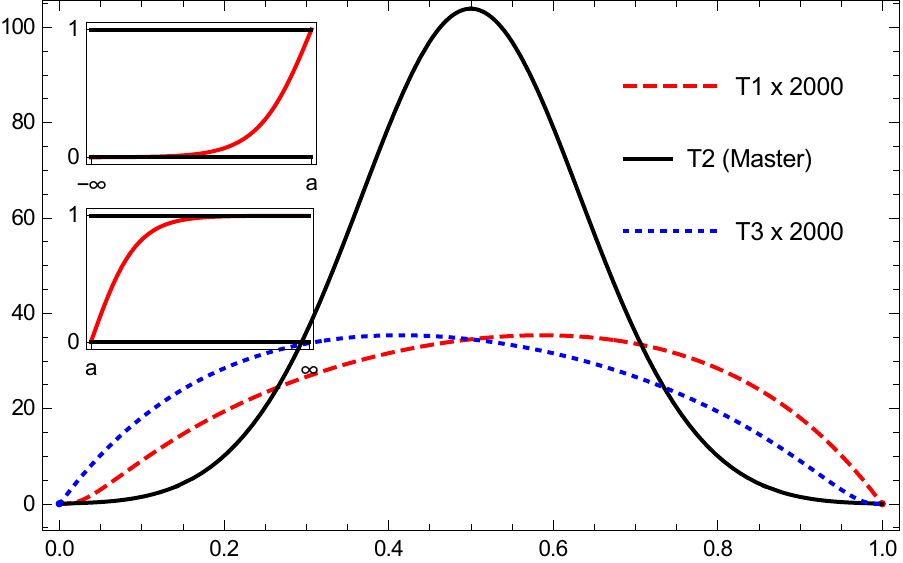}
}
\caption{Boltzmann factor $\text{exp}(-\text{Re}(S))$ vs.~flow
  parameter $\tau$ on the contributing thimbles in the U(1) one link
  model, where thimbles end in poles. Here it is necessary to
  transform integration ranges to maximize overlap. We chose to map
  all integration ranges to $\left[0,1\right]$, which can be done by a
  linear transformation for a thimble on $\left[a,b\right]$ or
  appropriate hyperbolic tangent transformations for intervals
  $\left[-\infty,a\right]$ or $\left[a,\infty\right]$, see inlays and
  \eqref{eq:trafo1} and \eqref{eq:trafo2} for the latter two
  transformations (we chose the free parameter of the transformation
  as $\xi=1.5$). For details see App.~\ref{app:integration}. This choice
  is not unique. Note that the curves have been rescaled for better
  visibility.}
\label{fig:dists}
\end{figure}
This model was studied by means of complex Langevin in
\cite{Aarts:2008rr}. Its thimble structure has been studied in
\cite{Aarts:2014nxa,Tanizaki:2017yow}. It has three different
contributing thimbles, of which two are connected by symmetry, see
Fig. \ref{fig:U1}. Looking at the Langevin drift in the complex plane,
one can see that there are two poles, in which all three thimbles
end. At those poles the drift diverges. The distributions after
mapping the integration on the finite interval $\left[0,1\right]$
according to App.~\ref{app:integration} are given in
Fig. \ref{fig:dists}.  Note that in this model, only periodic
observables which are analytic on U(1) make sense, hence we study the
analogue of the Polyakov loop, its inverse, the plaquette and the
density, which are given analytically in \cite{Aarts:2008rr}
\begin{align}
\nonumber \left<U\right>&=\left<e^{ix}\right>\,,\\[1ex]
\nonumber\left<U^{-1}\right>&=\left<e^{-ix}\right>\,,\\[1ex]
\nonumber\left<P\right>&=\left<\text{cos}(x)\right>\,,\\[1ex]
\left<n\right>&=\left<\frac{i\kappa \text{sin}(x-i\mu)}{1+\kappa \text{cos}(x-i\mu)}\right>\,.
\label{eq:observables}
\end{align}

\begin{table}[h]
\def\arraystretch{1.5}
\begin{tabular}{ |c | c | c |}
  \hline			
  $\mathcal{O}$ & numerical & exact \\

\hline\hline

Re$\left<U\right>$ & $0.315217(3)$ &$0.315219$\\ \hline
Re$\left<U^{-1}\right>$ & $1.800941(3) $&$1.800939$\\ \hline
Re$\left<P\right>$ &$1.058079(3) $&$1.058079$\\ \hline
Re$\left<n\right>$ &$0.742861(1)$ &$0.742860$\\ \hline
$\left. Z_2/Z_1\right|_{T_1} \times 10^{-3}$ &$2.99378(3)$ &$2.99382$\\ \hline
$\left.Z_1/Z_2\right|_{T_2} \times 10^{4}$ &$3.34032(4)$ &$3.34022$\\ \hline
$\left. Z_2/Z_3\right|_{T_3} \times 10^{-3}$ &$2.99377(3)$ &$2.99382$\\ \hline
$\left.Z_3/Z_2\right|_{T_2} \times 10^{4}$ &$3.34026(9)$ &$3.34022$\\ \hline
\end{tabular}
\caption{Numerical results and exact values of observables for the U(1) model with statistical errors. Note that the 
imaginary parts for the observables are all consistent with zero within the 
statistical error. Possible deviations are caused by numerical discretization errors, see main 
text for a detailed discussion.}
\label{tab:results}
\end{table}

Simulation results are given in Table \ref{tab:results}. For the simulations we have collected $\O(10^{9})$ measurements for the U(1) model. Again, errors have been estimated via a standard Jackknife analysis.

Our simulation results agree with the exact results from
\cite{Aarts:2014nxa}.  In Table \ref{tab:results} we only provide
statistical errors. Systematic errors arise from numerical
discretization artifacts along the thimble. In the simple cases at
hand the latter can be quantified by comparing the exact solution with
the result from integrating along the discretization obtained from the
gradient flow (\ref{eq:GradFlow}).  In the case of the observables
given in (\ref{eq:observables}) the deviation is of order
$10^{-6}$. Therefore the systematic error is comparable to the statistical error.  
The ratio of partition functions seems to be particularly sensitive to this effect. 
 However, by taking into account the statistical and the expected systematic
error all quantities agree with the exact result within the error.

\section{Conclusions and outlook}
\label{sec:concl}

We propose a new method based on Lefschetz thimbles for solving
theories with a sign problem. This method works with two steps: First
we find all contributing fixed points by scanning the original
manifold for intersecting anti-thimbles. We then project a grid of
points in the complex plane onto the thimbles -- which requires the
knowledge of the fixed points -- to obtain a numerical
parametrization. Second we simulate on these parametrizations and
determine the relative weights within the simulation by means of a
reweighting procedure. This reweighting can be tuned such that there
is no overlap problem. We remark that within this method finding
numerical parametrizations of the thimbles may be costly in higher
dimensions.  The reweighting procedure on the other hand
straightforwardly generalizes to field theories and can be combined
with other simulation algorithms for thimbles.

The above method has emerged from discussions and investigations of
the idea of simulating a complex Langevin evolution, that is directed
to the Lefschetz thimbles. This procedure of a Lefschetz-cooled Langevin
update has been partially, but not fully, successful. In our opinion
such a combined approach still has its potential, more details can be
found in App.~\ref{app:LCOOL}. 

Our proposal for simulating on Lefschetz thimbles has been illustrated
on a one-site $z^4$ model in Sec.~\ref{sec:modelz4}.  It is
successfully tested in a U(1) one-link model at finite density. The
results are discussed in detail in Sec.~\ref{sec:modelU1}.

Interesting future applications are field theories, e.g.~the Schwinger
model and higher-dimensional gauge theories, see
\cite{HD-Bielefeld}. 

\section*{Acknowledgements}
We thank K.~Fukushima, C.~Schmidt, A.~Rothkopf, F.~Ziesch\'{e}, the Heidelberg
Lattice group and the CLE collaboration for discussions and work on
related subjects. This work is supported by EMMI, the BMBF grant
05P12VHCTG, and is part of and supported by the DFG Collaborative
Research Centre "SFB 1225 (ISOQUANT)".  I.-O.~Stamatescu and
M.~Scherzer acknowledge financial support from DFG under
STA 283/16-2. F.P.G.~Ziegler is supported by the FAIR OCD
project.

\appendix

\section{Partition function weights}
\label{sec:deriv}
Here we give the derivation of (\ref{eq:ratio}).
\begin{align}
\nonumber \frac{Z_{1}^r}{Z_{2}^r}&= \frac{\int_{a}^{b} d\tau\, e^{-\text{Re}\left[S_1\left(\tau\right)\right]}}{\int_{a}^{b} d\tau\,e^{-\text{Re}\left[S_2\left(\tau\right)\right]}}\\[1ex]
\nonumber &= \frac{\int_{a}^{b} d\tau\, e^{-\text{Re}\left[S_1\left(\tau\right)+S_2\left(\tau\right)-S_2\left(\tau\right)\right]}}{\int_{a}^{b} d\tau\, e^{-\text{Re}\left[S_2\left(\tau\right)\right]}}\\[1ex]
\nonumber &= \frac{\int_{a}^{b} d\tau\, e^{-\text{Re}\left[S_2\left(\tau\right)\right]}e^{\text{Re}\left[S_2\left(\tau\right)-S_1\left(\tau\right)\right]}}{\int_{a}^{b} d\tau\, e^{-\text{Re}\left[S_2\left(\tau\right)\right]}}\\[1ex]
&=\left<e^{\text{Re}\left[S_2(\tau)-S_1(\tau)\right]}\right>_{2}^{r}\,.
\label{eq:derivation}
\end{align}
This derivation of the case given in (\ref{eq:derivation}) requires two presuppositions
\begin{itemize}
\item The flow parameters on both thimbles can be identified. We
  normalize the steepest descent equation in order to automatically
  fulfill this requirement. Note that for cases where different
  parametrizations occur, instead of one flow parameter $\tau$, there
  will be $\tau_1$ and $ \tau_2$ and the derivative $d\tau_1/d\tau_2$
  should be taken into account. However, for practical purposes it
  should be possible to normalize the steepest descent equations such
  that $\tau_1 = \tau_2$.
\item The integration boundaries are the same. This can be enforced easily by variable changes in 
the integral, see App.~\ref{app:integration}. 
Note that if the integration boundaries are the same from the beginning 
as for the case of (\ref{eq:derivation}), then there will be no overlap problem, since 
the fixed points give the main contribution on the thimbles and we chose our parametrization such 
that all fixed points correspond to $\tau = 0$. Hence the peaks of the distributions 
are at the same point.
\end{itemize}

\section{Mapping integration ranges}
\label{app:integration}
If different thimbles have different parameter ranges, one has to map
all of them to the same interval, here we choose the interval
$\left[0,1\right]$. In case of an integral in the range
$\left[a,b\right]$, a simple linear shift is enough.  For an integral
over $\tau\in\left[-\infty,a\right]$, one possible transformation is
\begin{equation}
x\rightarrow x^{\prime}=1+\text{tanh}\left(\xi\left(x-a\right)\right)\,,
\label{eq:trafo1}
\end{equation}
conversely for $\tau\in\left[a,\infty\right]$, the analogue is
\begin{equation}
x\rightarrow x^{\prime}=\text{tanh}\left(\xi\left(x-a\right)\right)\,,
\label{eq:trafo2}
\end{equation}
and for $\tau\in\left[-\infty,\infty\right]$ the mapping becomes
\begin{equation}
x\rightarrow x^{\prime}=\frac{1+\text{tanh}\left(\xi x\right)}{2}\,,
\label{eq:trafo3}
\end{equation}
where the parameter $\xi$ can be chosen such that the overlap of the
distributions in \eqref{eq:derivation} is maximal and hence the
overlap problem becomes small. The Jacobian of the transformation can
then be absorbed in the action for the Monte Carlo simulation. We
chose this transformation in the case of the U(1) one link model,
where we chose $\xi=1.5$. We only choose such transformations that
have sufficiently fast falloff at the boundaries such that those
regions are suppressed exponentially.

\section{Combining the Complex Langevin and Lefschetz thimble methods}
\label{app:LCOOL}
As there are many complicated steps for simulations on thimbles, it is
desirable to find simpler alternatives, which at best can be applied
blindly. One natural idea \cite{Nishimura:2017vav, Bluecher,
  Schlosser, Syrkowski, Unverzagt} is to combine the complex Langevin
evolution with the Lefschetz thimbles. A combination of both equations
is only consistent after an appropriate coordinate transformation. The
latter can adaptively be generated during the combined Langevin and
gradient flows. Due to its similarity to standard cooling algorithms
as well as the gauge cooling we call this process {\it Lefschetz cooling}. 

Despite its full success described below in particular for simple
Gau\ss ian models it is only partially successful in more complicated
models, notably already the $z^4$ model. While the method is not fully
successful yet, in our opinion it is still a very interesting one to
pursue. Its potential power is the self-adaptive {\it local} nature of
the simulation steps. However, this also poses the biggest conceptual
question: How does such a local procedure capture the global nature
of the intersection numbers $n_\sigma$ in
(\ref{eq:Thimble_basics}) correctly? Note that besides its formal
importance this question could be practically less important as it seems: in
most models under investigation so far we have $n_\sigma=1$.

Below, we list some ideas putting Lefschetz cooling to work and discuss
their viability and applications. While none of those proposed ideas
so far have managed to give quantitative correct results for
observables, they provide useful insight into possible realizations of
the approach.

\subsection{Variable transformations}
\label{app:trafos}
We aim to make the complex Langevin evolution compatible with
constraints characterizing the Lefschetz thimbles by means of variable
transformations.  The latter have been investigated in combination
with the complex Langevin evolution in a different context in
\cite{Aarts:2012ft}. There it was shown that the complex Langevin
evolution including a transformation can give correct results while
failing in the original formulation of the problem.  Here we pursue
the idea of having flow time-dependent variable transformations to
transform the complex Langevin evolution towards the thimbles of the
theory. This approach is natural in the sense that complex Langevin
should already be able to sample the relevant fixed points
\cite{Aarts:2014nxa}, note that this is in a similar spirit as for the
contraction algorithm \cite{Alexandru:2015xva}. By forcing the complex
Langevin evolution close to thimbles the sign problem should be
weakened and parameter regions that have been inaccessible so far may
be reached.  In the following discussion we consider again
one-dimensional integrals.  One rather general ansatz for such
variable transformation is the M\"obius transformation
\begin{equation}
z(u) := \frac{au+b}{cu+d}\,.
\label{eq:mobius}
\end{equation}
This rather general ansatz has four $\tau$-dependent parameters that
have to be determined during the simulation.  This turns out to be a
rather challenging task. We find that the transformation
(\ref{eq:mobius}) seems to introduce repulsive structures
destabilizing the evolution.  Hence, we focus on a special M\"obius
transformation, namely a rotation
\begin{equation}
z(u) := ue^{i\theta}\,, 
\label{eq:rotation}
\end{equation}
where $u$ takes the role of the (complex) field variable and $\theta$
is a $\tau$-dependent parameter.  Consider a point in the complex
plane sufficiently close to the thimble. Then, a rotation suffices to
map this point even closer to or onto the thimble. This indicates that
the transformation (\ref{eq:rotation}) is both necessary and
sufficient for fulfilling the constraints mentioned above.  For the
remaining part of this paper we always refer to the rotation
(\ref{eq:rotation}) when discussing variable transformations.

\subsection{Lefschetz cooling}
\label{app:Lcooling}
Thimbles are curves passing through the fixed points and along them
the imaginary part of the action is constant. This gives rise to
various constraints which we impose onto the complex Langevin
evolution by including the variable transformation
(\ref{eq:rotation}).

Let $u \in \mathbb{C}$ and $\theta \in \mathbb{R}$. The transformed action reads 
\begin{equation}
	 S_u := S(z(u)) - \log(z'(u))\,.
	\label{eq:actiontrafo}
\end{equation}
The procedure here is to be understood as a passive transformation,
see the appendix in \cite{Aarts:2012ft}.  Hence, the complex Langevin
equation in the transformed variables becomes
\begin{equation}
\partial_{\tau} u = -\frac{\partial S_u}{\partial u} + \eta\, ,
\label{eq:CLE-trafo}
\end{equation}
where $\eta \in \mathbb{R}$.

In the following we investigate how the $\tau$-dependent
transformation parameter $\theta$ evolves under the dynamics induced
by different constraints.
\subsubsection{Lefschetz cooling the transformed thimbles}
\label{app:ImSu}
First, we formulate the additional constraint completely in the
transformed theory \cite{Bluecher}, i.e.~we demand
\begin{align}
\mathrm{Im}(S_u) &= \text{const}\,.
\label{eq:constraint-u}
\end{align}
This constrains the evolution close to the thimbles in the transformed theory.
By taking the total $\tau$ derivative of this equation we obtain
\begin{equation}
\mathrm{Im}\left(\frac{\partial S_u}{\partial u} \dot{u}\right) 
+ \mathrm{Im}\left(\frac{\partial S_u}{\partial \theta} \right) \dot{\theta} = 0\,.
\label{eq:total-deriv-constraint}
\end{equation}
By inserting the Langevin evolution (\ref{eq:CLE-trafo}) 
we get the time evolution of the angle $\theta$
\begin{equation}
\frac{\partial \theta}{\partial \tau} = \frac{\mathrm{Im}\left(\left(\frac{\partial S_u}{\partial u}\right)^2\right)}{\mathrm{Im}\left(\frac{\partial S_u}{\partial \theta}\right)}\, .
\label{eq:theta-evolution}
\end{equation}

\subsubsection{Lefschetz cooling the original thimbles}
\label{app:ImSz}
Alternatively, we may formulate the constraint in the original theory which yields
\begin{equation}
\mathrm{Im}(S_z) = \text{const}\,.
\label{eq:constraint-z}
\end{equation}
Again taking the total derivative and inserting (\ref{eq:CLE-trafo}) 
the evolution for $\theta$ becomes
\begin{equation}
\frac{\partial \theta}{\partial \tau} 
= \frac{\mathrm{Im}\left(\frac{\partial S_z}{\partial z}\frac{\partial z}{\partial u}\frac{\partial S_u}{\partial u}\right)}{\mathrm{Im}\left(\frac{\partial S_z}{\partial z}\frac{\partial z}{\partial \theta}\right)}\, .
\label{eq:z-theta-evolution}
\end{equation}
\subsubsection{Applicability}
On both, the original and the transformed thimbles one can see that the denominator in the 
evolution equations for $\theta$ in (\ref{eq:theta-evolution}) and (\ref{eq:z-theta-evolution}) 
introduces poles. 

\begin{figure}
{
\centering
\includegraphics[width=0.9\columnwidth]{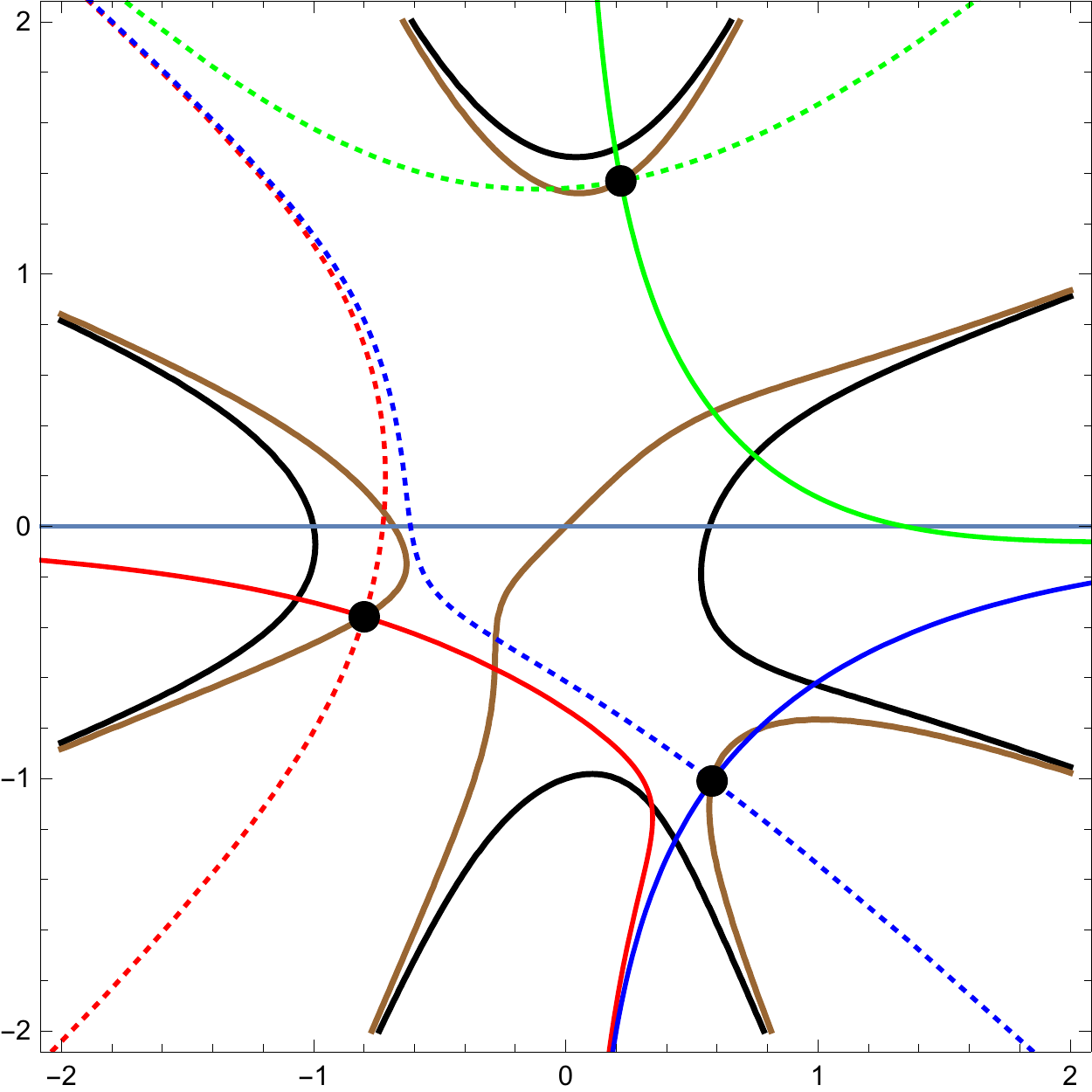}
}
\caption{Pole and thimble structure for the $z^4$ model enforcing the two constraints 
(\ref{eq:constraint-u}) and (\ref{eq:constraint-z}), each represented in the original variable $z$.
Red, green and blue solid (dashed) lines are the (anti-)thimbles and black points denote the fixed points. 
The brown solid lines are the poles from the constraint $\text{Im}[S_{z}]=\text{const}$, while the black 
solid lines represent the poles from the constraint $\text{Im}[S_{u}]=\text{const}$.}
\label{fig:poles}
\end{figure}

We analyze these in the representation of the original variable $z$ and show the denominators 
of (\ref{eq:theta-evolution}) and (\ref{eq:z-theta-evolution}) for the $z^4$ model 
respectively in Fig.~\ref{fig:poles}.

Those poles destabilize the numerical simulations and so far we have not found a way to resolve this. 
Our efforts for improvements include 
modifications to the noise term such that the poles are penalized, i.e.~one can 
multiply the noise by the denominator of \eqref{eq:theta-evolution} or  \eqref{eq:z-theta-evolution} 
respectively. 
However, this (i) leads to wrong expectation values and (ii) prohibits the Langevin evolution from 
jumping between the contributing thimbles. This prevents sampling of the correct thimble weights. 
In simple models such as the one-site Gau\ss ian model
\begin{equation}
S(z)=\frac{\sigma}{2}z^2\,,
\label{eq:Gauss}
\end{equation} (ii) is not a problem, since the model has only one thimble and (i) 
can be resolved by a linear rescaling. This is however due to the low complexity and high symmetry 
of the model and does not generalize.

\subsection{Advanced Lefschetz cooling}
\label{app:advLC}
Instead of explicitly demanding the thimble constraint to be fulfilled, 
it is possible to directly combine the complex Langevin evolution with the steepest descent equations.
\begin{align}
\frac{\partial u}{\partial \tau} &= -\frac{\partial S_u}{\partial u} + \eta \nonumber \\[1ex]
\frac{\partial u}{\partial \tau} &= -\overline{\frac{\partial S_u}{\partial u}}\,.
\label{eq:CLE-TH-flow}
\end{align}
Here the idea is that complex Langevin already takes into account all
relevant fixed points and the steepest descent equation keeps the
evolution close to the thimbles, alleviating the sign problem.  Taking
the difference of the previous two equations implies that the
imaginary part of $u$ remains constant which reduces the evolution in
$u$ to real Langevin. Hence we have
\begin{equation}
0 = \frac{\partial S_u}{\partial u} - \overline{\frac{\partial S_u}{\partial u}} 
\Rightarrow \partial_{\tau}\mathrm{Im}(u) = 0\,. 
\label{eq:const-im-u}
\end{equation}
This leads to the following evolution equation for $u$ and 
conditions to the angle $\theta$.
Thus we find by using the action as given in (\ref{eq:actiontrafo}) 
\begin{align}
  &\partial_{\tau} \mathrm{Re}\, u = - \mathrm{Re}\, \frac{
\partial S_u}{\partial u} + \eta \nonumber \\[1ex] 
  &\text{and} \nonumber\\[1ex]
  &\partial_{\tau}\text{Im}\, u=-\text{Im}\,\frac{\partial S_u}{\partial u}=
-\mathrm{Im}\left(\frac{\partial S}{\partial z} z^{\prime} 
    - \frac{z^{\prime \prime}}{z^{\prime}}\right) = 0\, .
\label{eq:real-Langevin-u}
\end{align}

Note that here we explicitly see, why a rotation should be
sufficient. We illustrate this by means of the Gau\ss ian model
(\ref{eq:Gauss}).  It holds that
$z^{\prime} = \mathrm{e}^{i\, \theta}$, $z^{\prime \prime} = 0$
leading to the following expression for the constraint in
(\ref{eq:real-Langevin-u})
\begin{align}
\mathrm{Im}(\sigma\, z\, z^{\prime})  = \mathrm{Im}(\sigma\, \mathrm{e}^{2i\theta} u) = 0\,.
\label{eq:gaussian-imag}
\end{align}
Here, $u\in\mathbb{R}$ and with $\sigma = \sigma_r\, \mathrm{e}^{i\theta_{\sigma}}$ the solution 
to (\ref{eq:gaussian-imag}) becomes 
\begin{equation}
\theta^{*} = - \frac{1}{2}\, \theta_{\sigma}\,.
\label{eq:cond-theta}
\end{equation}
This rotates the thimble in the original theory precisely onto the real axis of the transformed theory.

\begin{figure}[t]
{
\centering
}
\includegraphics[width=.9\columnwidth]{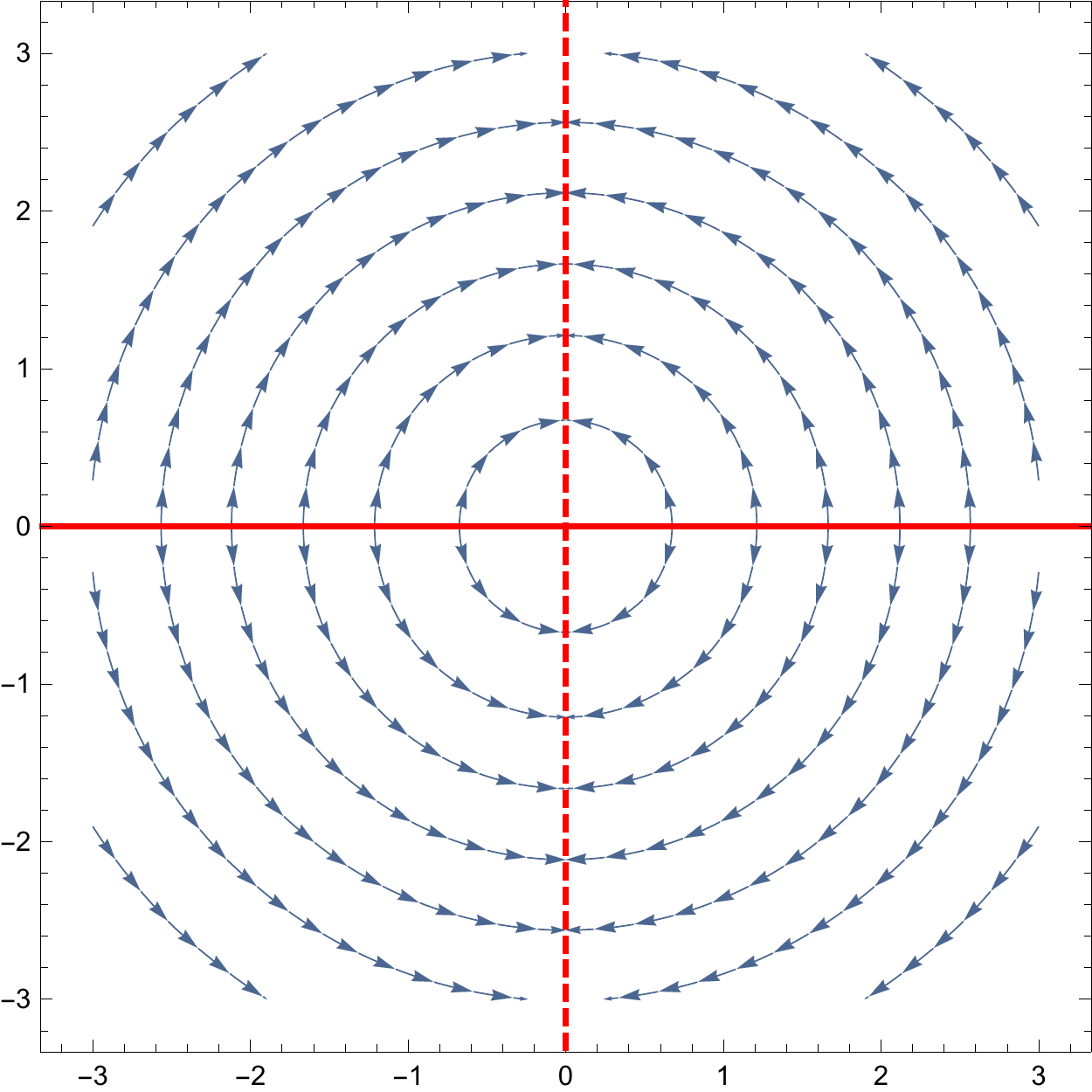}
\caption{Visualization of the rotation of the complex plane for the
  Gau\ss ian model with $\sigma=1+i$ according to
  \eqref{eq:evol-theta-Gauss} after rotating the thimble onto the real
  axis. The arrows point in the direction of rotation. Here, the
  transformed thimble is the real axis (solid red line, repulsive),
  while the anti-thimble is the imaginary axis (dashed red line,
  attractive).}
\label{fig:vis_sign}
\end{figure}

Next, we investigate how $\theta$ changes with the flow induced by (\ref{eq:const-im-u}).
Hence, we take the total $\tau$ derivative of (\ref{eq:const-im-u}), yielding 
\begin{equation}
\frac{d}{d\tau}\mathrm{Im}\left(\frac{\partial S_u}{\partial u}\right) = 0\,.
\label{eq:total-tau-deriv-imdriftu}
\end{equation}
This leads to
\begin{equation}
\mathrm{Im}\left(\frac{\partial^2 S_u}{\partial u^2} \dot{u}\right) 
+ \mathrm{Im}\left(\frac{\partial^2 S_u}{\partial u\, \partial \theta} \right) \dot{\theta} = 0\,,
\label{eq:m3-ode}
\end{equation}
from which we find 
\begin{equation}
\dot{\theta} = \frac{\mathrm{Im}\left(\dfrac{\partial^2 S_u}{\partial u^2}\, 
\dfrac{\partial S_u}{\partial u}\right)}{\mathrm{Im}\left(\dfrac{\partial^2 S_u}{\partial u\, \partial \theta} \right)}\, ,\\[2ex]
\label{eq:m3-theta}
\end{equation}
upon inserting the drift term for $\dot{u}$.
In this form, the dynamics are unstable. 
For instance in the Gau\ss ian model, where the (anti-)thimble is a straight line 
this manifests itself by a diverging evolution along the anti-thimble to infinity. 
\eqref{eq:cond-theta} rotates the thimble onto the real axis and transforms the anti-thimble 
into the imaginary axis.

Inserting the action of the Gau\ss ian model into \eqref{eq:m3-theta} we find 
\begin{equation}
	\dot \theta = \frac{1}{\sigma_r} \sin(\theta_{\sigma} + 2 \theta)\,.
	\label{eq:evol-theta-Gauss}
\end{equation}
Examining the numerical solution to the previous equation we easily see that the thimble is repulsive 
whereas the anti-thimble is attractive, see Fig.~\ref{fig:vis_sign} for an illustration represented in 
the transformed theory.

To render the thimbles attractive we consider again (\ref{eq:m3-ode}).
Replacing $\dot u$ by the Langevin drift with a reversed sign $+ \partial S_u /  \partial u$ reverses
the sign in (\ref{eq:m3-theta}) yielding
\begin{equation}
\dot{\theta} = -\frac{\mathrm{Im}\left(\dfrac{\partial^2 S_u}{\partial u^2}\, 
\dfrac{\partial S_u}{\partial u}\right)}{\mathrm{Im}\left(\dfrac{\partial^2 S_u}{\partial u\, \partial \theta} \right)}\,.
\label{eq:m3-theta-guess}
\end{equation}
This also enables and enforces hopping between the thimbles and 
inverts the stability properties of the fixed points. Note that the evolution
in $u$ is still governed by the complex Langevin equation \ref{eq:CLE-trafo}).

\begin{figure}[t]
{
\centering
  \includegraphics[width=0.9\columnwidth]{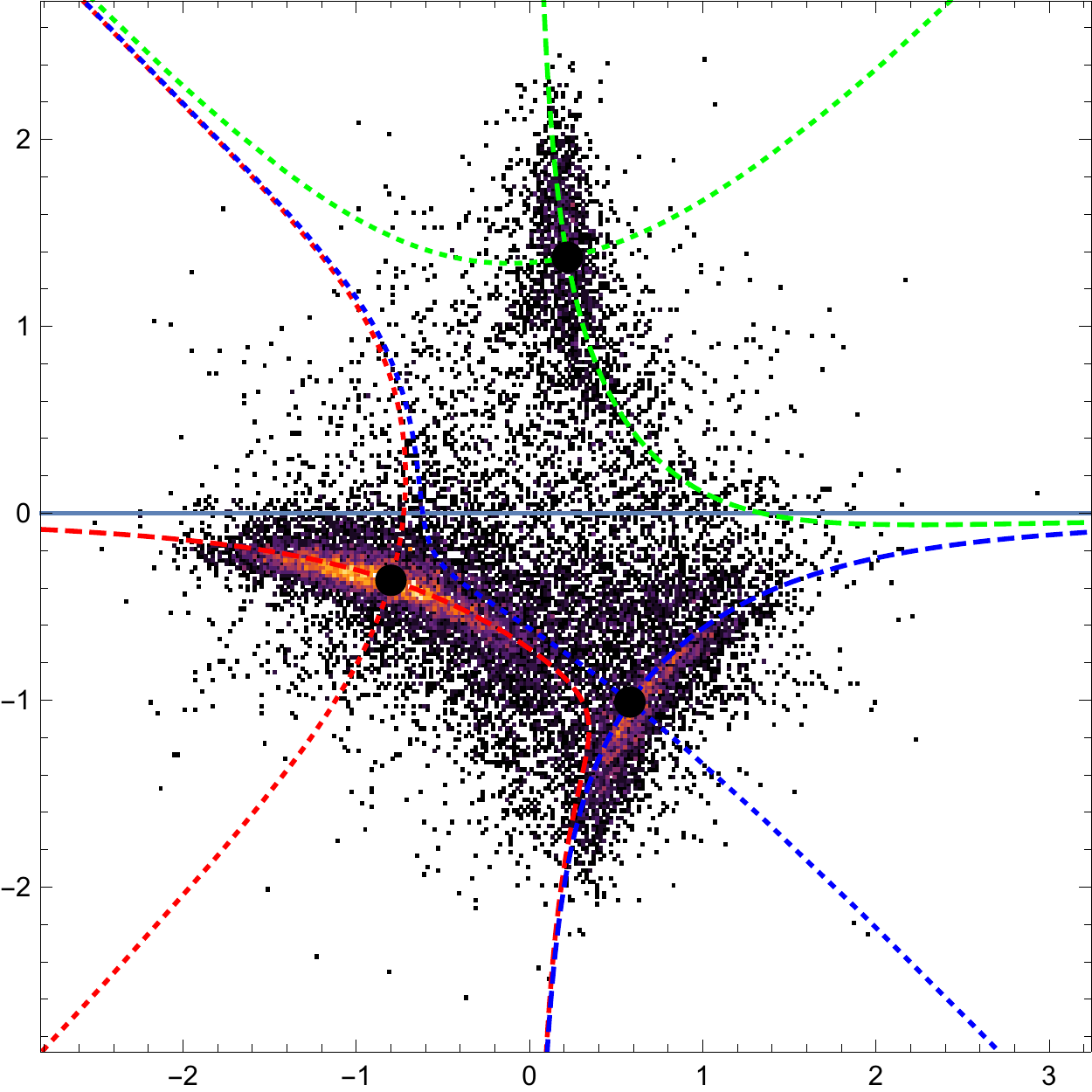}
  }
  \caption{Scatter plot of the combined complex Langevin evolution and
    thimble constraint dynamics in $\theta$ with the reversed sign
    mapping, see \eqref{eq:m3-theta-guess} and the discussion in the
    text. The proposed method is being applied to the $z^4$ model and
    the evolution is represented in the original variable $z$.  The
    scatter plot is color coded, where black corresponds to low
    density and yellow corresponds to high density. The mapping
    enforces the sampling of all relevant thimbles, as well as
    allowing for transitions between them and guarantees
    stability.}
\label{fig:opposite_sign}
\end{figure}

Therefore, inserting the sign-reversed drift for $u$ in
(\ref{eq:m3-theta}) can be understood in the sense of a mapping
with the following properties: it guarantees that the evolution stays
close to the thimbles, as well as allowing for transitions
between different contributing thimbles.  This approach yields the
correct result for the Gau\ss ian model.  However, once the thimble
structure becomes slightly more complicated, the values of observables
are not computed correctly. This has been tested for different actions
in \cite{Schlosser, Syrkowski}.  While the procedure samples all
thimbles, it does not correctly take into account their relative
weights.  Fig.~\ref{fig:opposite_sign} shows a scatter plot of the
$\tau$-evolution applied to the $z^4$ model. Clearly, this algorithm
samples all thimbles. But the contributing ones are being sampled with
a higher weight, see the yellow regions in the scatter plot.


\bibliography{literature}
\end{document}